\documentclass[aip,numerical,apl,showpacs,twocolumn,amsmath,amssymb,superscriptaddress]{revtex4}
\usepackage{graphicx}
\usepackage{bm}
\usepackage{amssymb}
\usepackage{amsmath}
\usepackage{latexsym}
\usepackage{color}
\newcommand{\indrm}[1]{_{\mathrm {#1}}}    
\newcommand{\trx}{1} 
\newcommand{\rtx}{2} 
\newcommand{\smx}{0} 
\newcommand{\thick}{d}    
\newcommand{\smxtal}{C$_{\indrm{\smx}}$}    
\newcommand{\rtxtal}{C$_{\indrm{\rtx}}$}   
\newcommand{\trxtal}{C$_{\indrm{\trx}}$}    
\newcommand{\sample}{S}    
\newcommand{\extlngth}{\Lambda}    
\begin{document}  
\title{Time-delayed beam splitting with energy separation of x-ray channels }
\author{Yuri P. Stetsko}
\author{Yuri V. Shvyd'ko }\email{shvydko@aps.anl.gov}
\author{G. Brian Stephenson}
\affiliation{Advanced Photon Source, Argonne National Laboratory,   Argonne, Illinois 60439, USA}

\begin{abstract} 
  We introduce a time-delayed beam splitting method based on the
  energy separation of x-ray photon beams.  It is implemented and
  theoretically substantiated on an example of an x-ray optical scheme
  similar to that of the classical Michelson interferometer. The
  splitter/mixer uses Bragg-case diffraction from a thin diamond
  crystal. Another two diamond crystals are used as
  back-reflectors. For energy separation the back-reflectors are set
  at slightly different temperatures and angular deviations from exact
  backscattering. Because of energy separation and a minimal number
  (three) of optical elements, the split-delay line has high
  efficiency and is simple to operate. Due to the high transparency of
  diamond crystal, the split-delay line can be used in a beam sharing
  mode at x-ray free-electron laser facilities. The delay line can be
  made more compact by adding a fourth crystal.
\end{abstract}



\maketitle

Complex nanoscale dynamics in condensed matter can be studied in a
broad dynamic range by x-ray photon correlation spectroscopy (XPCS)
using coherent x-rays from x-ray free-electron lasers (XFELs)
\cite{GSG07}.  The split-pulse technique is one of the promising
approaches to access dynamics from femtosecond to nanosecond
regimes. In this technique, each x-ray pulse is split into reference
and delayed pulses of equal intensity, arriving at the sample separated
in time.  The scattering from the reference and delayed pulses is then
collected during the same exposure of an area detector.

The principle optical scheme of the split-pulse technique was
discussed in \cite{Tesla-TDR,Altarelli:1088597,GSG07}. The main
components are a splitter and mixer crystals in Bragg diffraction, and
two additional Bragg reflecting crystals guiding the delayed pulse and
controlling the delay. To ensure very short (femtosecond length) and
even negative delays, an advanced scheme was proposed and realized
\cite{LCLS,RFS09,RFS11}. It has  additional four crystals (total
eight) designed to guide the reference pulse. The large number of
optical elements inevitably complicates alignment and operations, and
also compromises the split-delay line efficiency.

\begin{figure}[t!]
\setlength{\unitlength}{\textwidth}
\begin{picture}(1,0.65)(0,0)
\put(0.0,0.0){\includegraphics[width=0.5\textwidth]{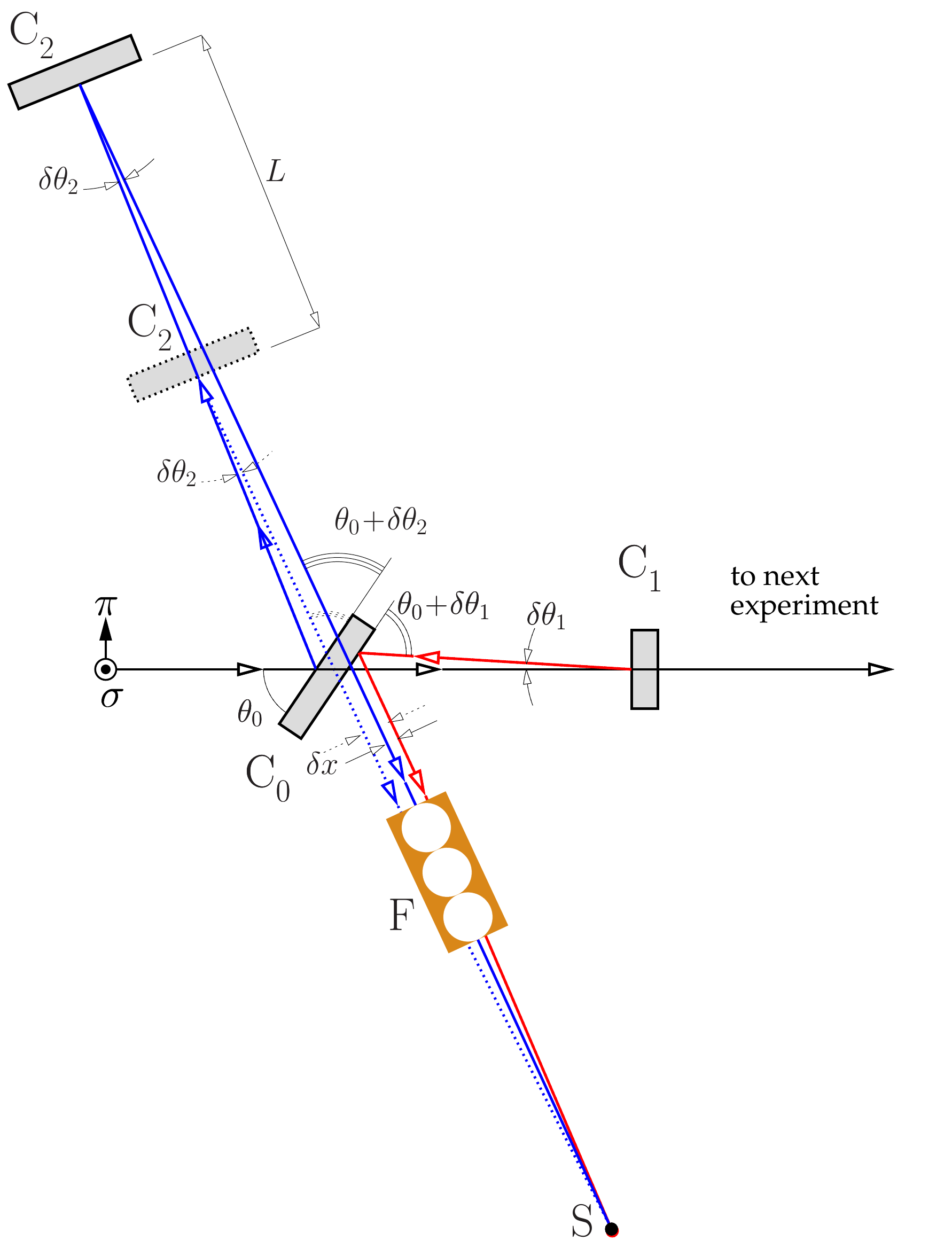}}
\end{picture}
\caption{Optical scheme of a three-crystal \smxtal , \trxtal , \rtxtal ,  split-delay line in a
  Michelson-interferometer-type configuration using dual-energy
  splitting (see text for details).  
  The reference (red) and delayed (blue) beams are shifted parallel to
  each other by $\delta x $, provided the angular deviations from
  exact backscattering $\delta\theta_{\indrm{\trx}}=
  \delta\theta_{\indrm{\rtx}}$. The parallel beams are focused on
  sample S by focusing system F.  The shift $\delta x$, see
  Eq.~\eqref{eq010}, is due to a finite thickness $\thick$ of crystal
  \smxtal , and a delay path length $L$ determining the nonzero x-ray
  pulse delay $\tau=2L/c$.  }
\label{fig001}
\end{figure}

Here we introduce and study theoretically the performance of a
split-delay x-ray scheme with the minimal amount of crystals, three,
as shown in Fig.~\ref{fig001}. The scheme resembles the classical
Michelson interferometer \footnote{We note here that an x-ray version
  of the Michelson interferometer, designed for hard x-ray Fourier
  spectroscopy applications, was already implemented in hard x-ray
  optics; however, it used many (eight) crystals and Bragg reflections
  in far from exact backscattering conditions
  \cite{AB91,NB03,SIK04}.}. Crystal \smxtal\ plays the roles of both
the splitter and the mixer. Crystals \rtxtal\ and \trxtal\ are set to
reflect x-rays in almost exact Bragg backscattering geometry, and
guide the reference (red) and delayed (blue) pulses to sample \sample\
through \smxtal\ or by reflecting from \smxtal , respectively. The
delay $\tau=2L/c$ is varied by changing the crystal \rtxtal\ spatial
position along the beam, i.e., a delay path $L$.

The central problem in the design of a split-delay line is how to
split the incident beam and how to bring together the reference and
delayed beams on the sample.  In the traditional scheme, which we term
as single-energy splitting (SES), photons of the same energy are
divided between the reference and delayed beams.  Equal intensity
splitting can be realized by using Bragg diffraction in the reflection
(Bragg-case) geometry. To transmit 50\% of the incident beam into the
forward direction, crystal thickness should be close to the
extinction length $\extlngth $. For suitable Bragg reflections in
silicon, diamond, and some other crystals, $\extlngth \approx
1-10~\mu$m, which requires very thin crystals, the manufacture of which
is a challenging but now feasible task \cite{OYS13}.
Alternatively, Bragg diffraction in the transmission (Laue-case)
geometry can be used. The thicknesses of the splitter has to be chosen
as an integer of the Pendell\"osung length, which is closely related
to $\extlngth $ \cite{Authier}. Also in this case, equal intensity
splitting can be realized; however, one should be aware of the
following potential problems. First, the intensity between the two
channels is very sensitive to thickness errors. Second, the angular
acceptance of the splitter as well as the beams intensities
sufficiently decrease with increasing crystal thickness.

\begin{figure}[t!]
\setlength{\unitlength}{\textwidth}
\begin{picture}(1,0.56)(0,0)
\put(-0.15,-0.33){\includegraphics[width=0.80\textwidth]{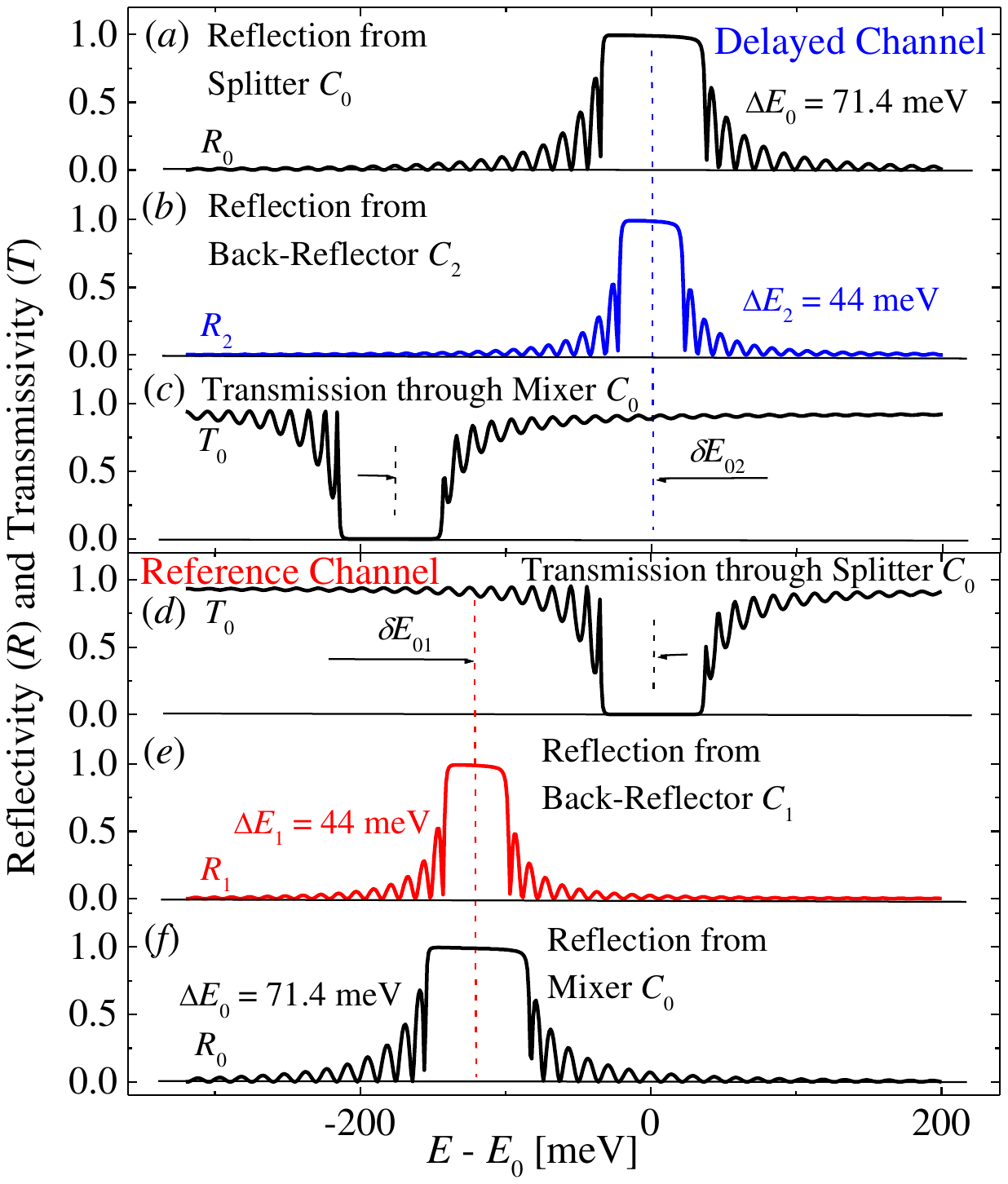}}
\end{picture}
\caption{Spectral dependencies of reflection (R) from and transmission
  (T) through individual crystals of the split-delay line, calculated
  in the framework of the dynamical theory of x-ray diffraction, with
  crystal parameters of device 1, given in Table~\ref{tab1}.}
\label{fig002}
\end{figure}

In the present Letter we study a beam splitting method based on the
photon energy separation of x-ray beams, which we refer to here as
dual-energy splitting (DES) \footnote{While working on the manuscript,
  it came to our attention that a similar idea of energy separation
  of the x-ray beams has been discussed in \cite{OYS13} for a very thin
  splitter without performing theoretical analysis or experiments. }.
DES is implemented here and studied theoretically on the example of
the x-ray Michelson-interferometer-type optical scheme, as shown in
Fig.~\ref{fig001}, and briefly presented above.  A few more details
regarding the scheme are in order.

The splitter/mixer crystal \smxtal\ reflects x-ray photons in an
energy bandwidth $\Delta E_{\indrm{\smx}}$ into the delayed pulse
channel, shown in blue in Fig.~\ref{fig001}. The reflectivity is close
to 100\% if the crystal thickness $\thick \gg \extlngth$, see
Fig.~\ref{fig002}(a). The incidence angle $\theta_{\indrm{\smx}}$ is
such that the central photon energy $E_{\indrm{\smx}}$ of the spectral
distribution coincides with the central energy $E_{\indrm{\rtx}}$ of
the reflection bandwidth from crystal \rtxtal\ set into
backscattering, see Fig.~\ref{fig002}(b).  The bandwidth $\Delta
E_{\indrm{\rtx}}$ of crystal \rtxtal\ is chosen to be close to the
bandwidth $\Delta E_{\indrm{\smx}}$ of crystal \smxtal . The x-rays
back-reflected from \rtxtal\ are transmitted through \smxtal\ and
steered onto the sample \sample , provided the angular offset
$\delta\theta_{\indrm{\rtx}}$ is larger than the angular width
$\Delta\theta_{\indrm{\smx}} = ({\Delta
  E_{\indrm{\smx}}}/{E_{\indrm{\smx}}}) \tan\theta_{\indrm{\smx}}$ of
the Bragg reflection from \smxtal .  This can be easily realized,
since the angular width of back-reflection
$\Delta\theta_{\indrm{\rtx}} = 2\sqrt{\Delta
  E_{\indrm{\rtx}}/E_{\indrm{\rtx}}}$ is much larger \cite{KM72,
  Shvydko-SB}.

X-ray photons with energies outside the energy bandwidth $\Delta
E_{\indrm{\smx}}$ are transmitted through \smxtal , see
Fig.~\ref{fig002}(d), and guided into the reference pulse channel by
back reflection from \trxtal , as shown in red in Fig.~\ref{fig001}.
Crystal \trxtal\ is equivalent to \rtxtal . However, it is maintained
at a different temperature $T_{\indrm{\trx}} = T_{\indrm{\rtx}}+\delta
T_{\indrm{\trx\rtx}}$ to reflect x-rays transmitted through \smxtal\
in a bandwidth centered at $E_{\indrm{\trx}}$, which is shifted from
$E_{\indrm{\rtx}}$ by $\delta
E_{\indrm{\smx\trx}}=E_{\indrm{\smx}}-E_{\indrm{\trx}}>(\Delta
E_{\indrm{\smx}} + \Delta E_{\indrm{\trx}})/2$, see
Fig.~\ref{fig002}(e).  Photons back-reflected from \trxtal\ will be
reflected from \smxtal , see Fig.~\ref{fig002}(f), and directed onto
the sample, provided the angle of incidence to \smxtal\ is
$\theta_{\indrm{\smx}}+\delta\theta_{\indrm{\trx}}$, where according
to Bragg's law $\delta\theta_{\indrm{\trx}} = ({\delta
  E_{\indrm{\smx\trx}}}/{E_{\indrm{\smx}}})
\tan\theta_{\indrm{\smx}}$.  The same angular deviation
$\delta\theta_{\indrm{\trx}}$ is required in backscattering from
\trxtal , see Fig.~\ref{fig001}.

If $\delta\theta_{\indrm{\trx}}=\delta\theta_{\indrm{\rtx}}$, the two
beams propagate to the sample parallel to each other with a small
offset
\begin{equation}
\delta x\,=\,\ 2\thick_{\indrm{\smx}}\cos\theta_{\indrm{\smx}}\,-\,L\, \delta\theta_{\indrm{\trx}}.
\label{eq010}
\end{equation}
For $d_{\indrm{0}}=50~\mu$m the first term in Eq.~\eqref{eq010} is
about $60~\mu$m, while the second term varies from zero for $L=0$ to
$-30~\mu$m for $L=1.5$~m. As a result, $\delta x$ varies from
$60~\mu$m to $30~\mu$m, respectively. Given, a usual XFEL beam size of
$\simeq 300-700~\mu$m, such shift is insignificant. Despite the shift,
focusing system F brings all the photons to the same point on the
sample.

\begin{table*}[t!]
\centering
\begin{tabular}{|c|c||l|l|l|l|l|l|l|l|l||l|l|}
  \hline 
device &  polari- & crystal & material        & $(hkl)$ & $T_{\indrm{n}}$  & $E_{\indrm{n}}$      & $\Delta E_{\indrm{n}}$ & $\theta_{\indrm{n}} $         & $\Delta \theta_{\indrm{n}} $ &  $d_{\indrm{n}}$ & $\varepsilon_{\indrm{n}}$ & $\varepsilon_{\indrm{n}}^{{\mathrm (abs)}}$ \\[0pt]    
 \# & zation & C$_{\indrm{n}}$       &                 &         & [K]  & [keV]    & [meV]      &              & [$\mu$rad]       &  [$\mu$m] & \% & \% \\[0pt]    
  \hline   \hline    \hline  
 & & C$_{\indrm{\smx}}$   & diamond &  (004)  & 300  & 8.51389  & 72       & 54.7359$^{\circ}$  & 12             & 50   & -- & -- \\[-0.0pt]
\cline{3-13}
1 & $\sigma$ & C$_{\indrm{\trx}}$ & diamond &  (224)  & $T_{\indrm{\rtx}}+\delta T_{\indrm{\trx\rtx}}$ & $E_{\indrm{\smx}}-\delta E_{\indrm{\smx\trx}}$  & 44       & 90$^{\circ}$
-$\delta\theta_{\indrm{\trx}}\!/2$ & 3675             & 50   & 94 & 9.9 \\[-0.5pt]
 & &   &         &         & $\delta T_{\indrm{\trx\rtx}}$=13~K & $\delta E_{\indrm{\smx\trx}}$=120~meV  &            & $\delta\theta_{\indrm{\trx}}$=20~$\mu$rad &                  &      & & \\[-0.0pt]
\cline{3-13}
 & & C$_{\indrm{\rtx}}$   & diamond &  (224)  & 300  & $E_{\indrm{\smx}}$  & 44    & 90$^{\circ}$-$\delta\theta_{\indrm{\rtx}}\!/2$ & 3675             & 50   & 93 & 9.8 \\[-0.0pt]
\cline{3-13}
 & & C$_{\indrm{3}}$   & diamond &  (220)  & 300  & $E_{\indrm{\smx}}$  & 176    & 90$^{\circ}-\theta_{\indrm{\smx}}$ & 15             & $>$50   & -- & -- \\[-0.0pt]
  \hline   \hline    \hline  
 & & C$_{\indrm{\smx}}$   & diamond &  (220)  & 300  & 8.51389  & 66       & 35.2641$^{\circ}$  & 5.5             & 50   & -- & -- \\[-0.0pt]
\cline{3-13}
2 & $\pi$ & C$_{\indrm{\trx}}$ & diamond &  (224)  & $T_{\indrm{\rtx}}+\delta T_{\indrm{\trx\rtx}}$ & $E_{\indrm{\smx}}-\delta E_{\indrm{\smx\trx}}$  & 44       & 90$^{\circ}$
-$\delta\theta_{\indrm{\trx}}\!/2$ & 3675             & 50   & 85 & 8.6 \\[-0.5pt]
 & &   &         &         & $\delta T_{\indrm{\trx\rtx}}$=13~K & $\delta E_{\indrm{\smx\trx}}$=120~meV  &            & $\delta\theta_{\indrm{\trx}}$=10~$\mu$rad &                  &      & & \\[-0.0pt]
\cline{3-13}
 & & C$_{\indrm{\rtx}}$   & diamond &  (224)  & 300  & $E_{\indrm{\smx}}$  & 44    & 90$^{\circ}$-$\delta\theta_{\indrm{\rtx}}\!/2$ & 3675    & 50   & 84 & 8.5 \\[-0.0pt]
  \hline   \hline  
  \hline 
\end{tabular}
\caption{Crystal elements C$_{\indrm{n}}$ ($n=\smx,\trx,\rtx$) of two  split-delay lines schematically presented in Fig.~\ref{fig001}, and
C$_{\indrm{n}}$ ($n=\smx,\trx,\rtx, 3$) of the  split-delay lines schematically presented in Fig.~\ref{fig004}.  
  Crystal, Bragg reflection parameters, and incident radiation polarization states are given as used in all dynamical theory calculations:
  $(hkl)$ - Miller indices of Bragg reflections; $T_{\indrm{n}}$ - crystal temperature, $d_{\indrm{n}}$ - crystal thickness, 
  $\theta_{\indrm{n}}$ - glancing angle of incidence; $E_{\indrm{n}}$ - photon energy at the reflection curve center; 
  $\Delta E_{\indrm{n}}$, and $\Delta \theta_{\indrm{n}}$  are Bragg's reflection spectral width, and angular acceptance, respectively.  
  $\varepsilon_{\indrm{n}}$ and $\varepsilon_{\indrm{n}}^{{\mathrm (abs)}}$ are the relative and absolute efficiencies of respective channels.}     
\label{tab1}
\end{table*}

\begin{figure}[t!]
\setlength{\unitlength}{\textwidth}
\begin{picture}(1,0.50)(0,0)
\put(-0.15,-0.41){\includegraphics[width=0.80\textwidth]{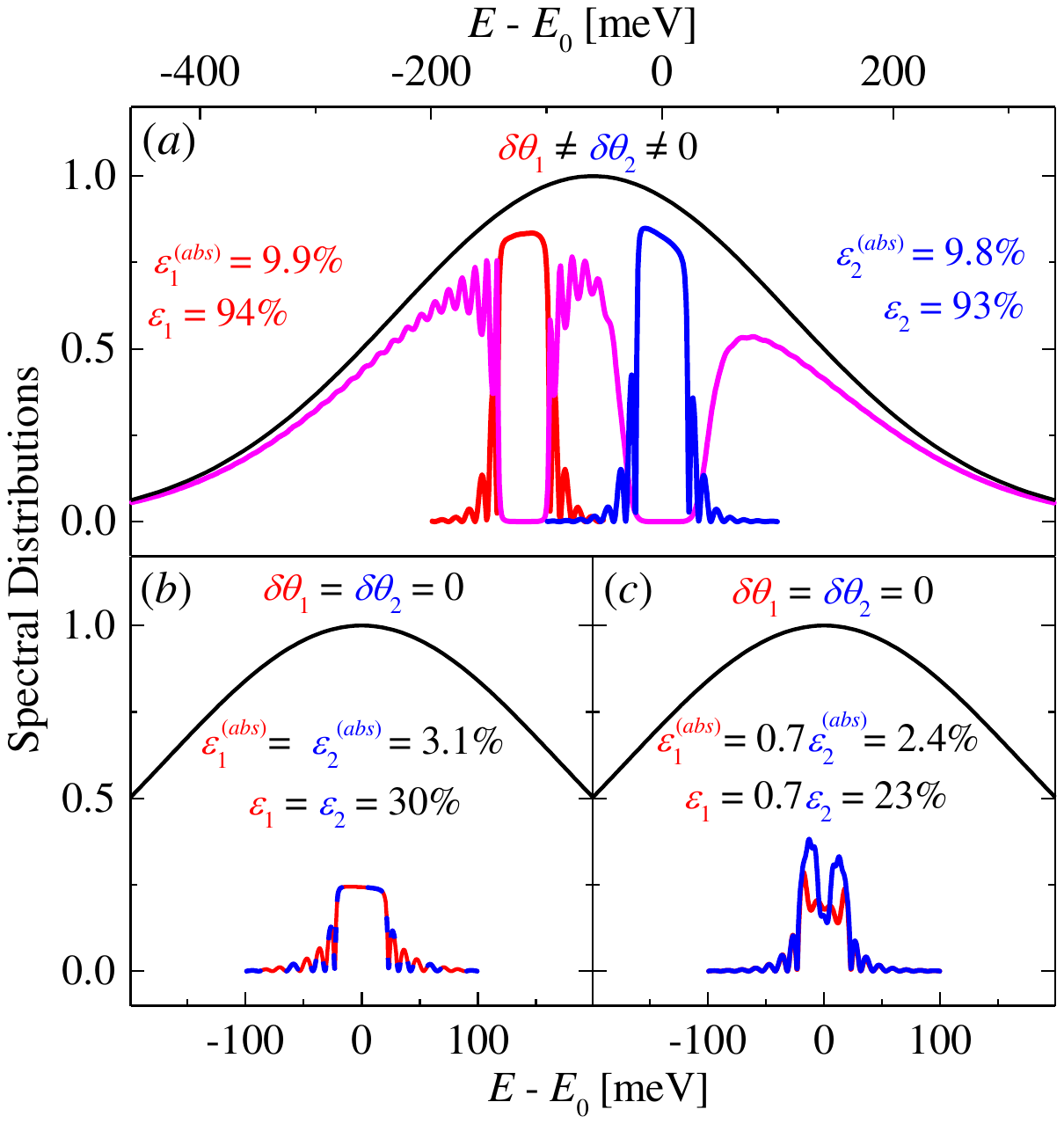}}
\end{picture}
\caption{Spectral distribution of x-rays arriving on the sample in the
  reference (red line) and delayed (blue line) channels, calculated
  using the dynamical theory of x-ray diffraction, with crystal parameters
  of device 1, given in Table~\ref{tab1}. Black line shows the
  spectrum of incident x-rays from a seeded XFEL.  The magenta line
  shows the spectral distribution of the photons, transmitted through
  crystals \smxtal\  and \trxtal . (a) Calculations for the
  dual-energy splitting case.  (b)-(c) Calculations under similar
  conditions, but with \smxtal\ functioning in the single-energy
  splitting mode in Bragg-case and Laue-case diffractions,
  respectively.}
\label{fig003}
\end{figure}


Spectral dependencies of reflection from and transmission through each
crystal optical element of the split-delay line are presented in
Fig.~\ref{fig002}, numerically calculated according to \cite{SC97} 
using equations of the dynamical theory of x-ray diffraction. Spectral
distribution of x-rays arriving on the sample in the reference and
delayed channels are shown in Fig.~\ref{fig003}(a) by the red and blue
lines, respectively. The calculations are performed assuming that the
XFEL is working in a self-seeding mode providing x-rays in a $400$-meV
bandwidth (black line) \cite{HXRSS12}, and with an angular spread of
$2.5~\mu$rad (FWHM).

The calculations presented in Figs.~\ref{fig002} and \ref{fig003}, are
performed with crystal parameters given in Table~\ref{tab1} for device
1.  The choice of the crystals is not unique. In this particular
example, diamond crystals are chosen for all three optical elements
for several reasons.  First, the photoabsorption in diamond is much
less than in Si, and therefore the efficiency of diamond optics is
higher. Second, for the same photon energy, the spectral bandwidth of
Bragg back-reflections from diamond crystals is larger, due to larger
Debye-Waller factors (larger Debye temperature) \cite{SSC10,SSB11}.
We have chosen Bragg reflections with the largest bandwidth and
therefore with the highest efficiency, applicable in a comfortable for
XPCS experiments photon range of 8-9~keV.  Given the very recent
advancement in fabrication of  high-quality diamond crystals and
their use in high-resolution, low-loss x-ray optics
\cite{BCC09,PDK11,SSB11,HXRSS12,SBT13}, the proposed configuration with
diamond crystal elements is deemed to be feasible.

Bragg back-reflection is very often accompanied by parasitic Bragg
reflections \cite{Sutter01,Shvydko-SB,LeSh04,CSC08}, which may waste a
significant number of useful photons and reduce the optics
efficiency. The choice of back-reflection was actually dictated also
by the requirement of the minimal amount of the parasitic reflections.
In particular, the 422  back-reflection from \trxtal\ and
\rtxtal\ diamond crystals considered here is accompanied only by one
pair of the parasitic Bragg reflections: 400 and 022. It is easy to
suppress them. To do this, the crystal plane containing the (422),
(400), and (022) reciprocal vectors has to be inclined by $\simeq
100~\mu$rad to the 422 Bragg diffraction plane.

Efficiency is a figure of merit for a split-delay line.  Relative
spectral efficiencies for the reference and delayed channels are very
high: $\varepsilon_{\indrm{\trx}}\simeq 94$\% and
$\varepsilon_{\indrm{\rtx}}\simeq 93$\%, respectively. The relative
spectral efficiency is defined as the number of photons in the
relevant channel normalized to the number of incident photons within
the channel bandwidth. Absolute spectral efficiencies are calculated
by normalization to the total number of incident photons.  For the
400-meV bandwidth of the XFEL radiation, they are equal to
$\varepsilon_{\indrm{\trx}}^{{\mathrm (abs)}}=9.9$\% and
$\varepsilon_{\indrm{\rtx}}^{{\mathrm (abs)}}=9.8$\%, respectively.
The total absolute efficiency of the scheme is about 20\%.  The rest
of the photons, $\simeq 60$\%, transmitted through both \smxtal\ and
\trxtal\ can be utilized by the downstream experiment. Therefore, such
a split-delay line can be used in a beam sharing mode at
XFEL facilities.

\begin{figure}[t!]
\setlength{\unitlength}{\textwidth}
\begin{picture}(1,0.45)(0,0)
\put(0.0,0.0){\includegraphics[width=0.5\textwidth]{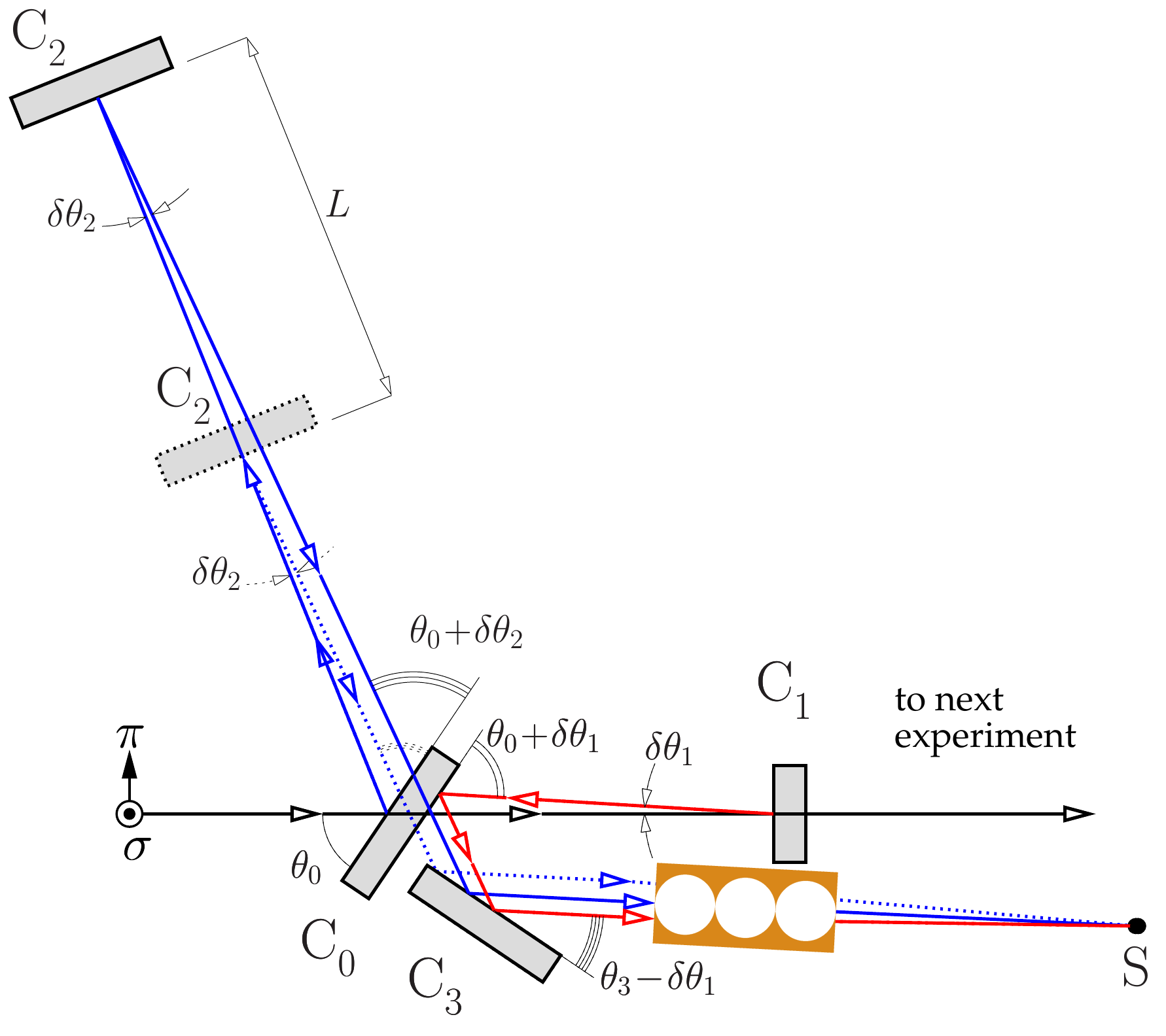}}
\end{picture}
\caption{By adding crystal C$_{\indrm{3}}$ with
  $\theta_{\indrm{3}}=\pi/2-\theta_{\indrm{\smx}}$ the delay line in
  Fig.~\ref{fig001} becomes more compact, with reference and delayed
  pulses propagating almost parallel to the incident beam. }
\label{fig004}
\end{figure}

Figures~\ref{fig003}(b) and \ref{fig003}(c) show for comparison
the results of calculations of the spectral distributions for the same
scheme, but, with \smxtal\ functioning in single-energy splitting
mode in Bragg-case (b) or Laue-case (c) diffraction, with
$\thick_{\indrm{\smx}}=6.4~\mu$m and $\thick_{\indrm{\smx}}=52.4~\mu$m,
respectively. Single-energy splitting requires that
$\delta\theta_{\indrm{\rtx}}=\delta\theta_{\indrm{\trx}}=0$.
The values of efficiencies indicated in Fig.~\ref{fig003}  show that the
dual-energy splitting scheme is about three to four times more
efficient than the single-energy splitting schemes.

If the vertical configurationis inconvenient, the same scheme could be
used in the horizontal scattering geometry.  In this case the incident
radiation is in the $\pi$-polarization state.  Device 2 in
Table~\ref{tab1} presents an example of the split-delay line
functioning in horizontal scattering plane and having very similar
performance in terms of efficiency and other parameters.

The scheme could be made more compact, with the beams propagating to
the sample parallel to the incident beam.  This is achieved by
additional Bragg reflection from an additional crystal C$_{\indrm{3}}$
with Bragg angle $\theta_{\indrm{3}}=\pi/2-\theta_{\indrm{\smx}}$, as
show in Fig.~\ref{fig004}.  The efficiency of the four-crystal scheme
is almost the same as that of the three-crystal scheme, provided a
high-reflectivity diamond crystal C$_{\indrm{3}}$ is used, as
suggested in Table~\ref{tab1}.

The split-delay lines presented here are also applicable at synchrotron radiation
facilities. The larger angular divergence of the incident beam $\simeq
10-15~\mu$rad will, however, result in a $\simeq 15-30$\% reduction of
the efficiency for the three-crystal scheme and a $\simeq 20-35$\%
reduction of the efficiency for the four-crystal scheme.
 
In conclusion, the three-crystal split-delay x-ray scheme for XPCS
applications is introduced and studied theoretically. Application of
the dual-energy splitting significantly increases the efficiency of
the optical scheme. Due to the high transparency of diamond crystal,
the split-delay line of such design can be used at XFEL facilities in the
beam sharing mode. A four-crystal modification makes the scheme
in-line and more convenient for XPCS experiments.

Work was supported by the U.S. Department of Energy, Office of
Science, under Contract No. DE-AC02-06CH11357.


\end{document}